\documentclass[fleqn,usenatbib]{mnras}

\usepackage{newtxtext,newtxmath}
\usepackage[T1]{fontenc}

\usepackage{amsmath}
\usepackage{stfloats}
\usepackage{graphicx}	

\newcommand{\msun}{\hbox{M$_{\odot}$}}

\title[A TESS View of ASASSN-18el]{TESS Shines Light on the Origin of the Ambiguous Nuclear Transient ASASSN-18el}

\author[Hinkle et al.]
{\href{http://orcid.org/0000-0001-9668-2920}{Jason T. Hinkle}$^{1}$\thanks{jhinkle6@hawaii.edu},
\href{https://orcid.org/0000-0001-6017-2961}{Christopher S. Kochanek}$^{2,3}$,
\href{https://orcid.org/0000-0003-4631-1149}{Benjamin J. Shappee}$^{1}$, 
\href{https://orcid.org/0000-0001-5661-7155}{Patrick J. Vallely}$^{2}$,
\newauthor
\href{https://orcid.org/0000-0002-4449-9152}{Katie Auchettl}$^{4,5,6}$, 
\href{https://orcid.org/0000-0002-9113-7162}{Michael Fausnaugh}$^{7}$, 
\href{http://orcid.org/0000-0001-9206-3460}{Thomas W.-S. Holoien}$^{8}$\thanks{NHFP Einstein Fellow}, 
\href{https://orcid.org/0000-0003-0660-9776}{Helena P. Treiber}$^{1,9}$, 
\newauthor
\href{https://orcid.org/0000-0003-3490-3243}{Anna V. Payne}$^{1}$\thanks{NASA Fellowship Activity Fellow},
{B. Scott Gaudi}$^{2}$,
\href{https://orcid.org/0000-0002-3481-9052}{Keivan G. Stassun}$^{10}$,
\href{http://orcid.org/0000-0003-2377-9574}{Todd A.~Thompson}$^{2,3}$,
J. L. Tonry$^{1}$,
\newauthor
and \href{http://orcid.org/0000-0001-6213-8804}{Steven Villanueva Jr.}$^{11}$\thanks{NPP Fellow} \\
$^{1}$Institute for Astronomy, University of Hawai`i, 2680 Woodlawn Dr., Honolulu, HI 96822, USA\\
$^{2}$Department of Astronomy, The Ohio State University, 140 West 18th Avenue, Columbus, OH 43210, USA\\
$^{3}$Center for Cosmology and Astroparticle Physics, The Ohio State University, 191 W.~Woodruff Avenue, Columbus, OH 43210, USA\\
$^{4}$School of Physics, The University of Melbourne, Parkville, VIC 3010, Australia\\
$^{5}$ARC Centre of Excellence for All Sky Astrophysics in 3 Dimensions (ASTRO 3D), Australia\\
$^{6}$Department of Astronomy and Astrophysics, University of California, Santa Cruz, CA 95064, USA\\
$^{7}$Department of Physics and Kavli Institute for Astrophysics and Space Research, Massachusetts Institute of Technology, Cambridge, MA 02139, USA\\
$^{8}$The Observatories of the Carnegie Institution for Science, 813 Santa Barbara St., Pasadena, CA 91101, USA\\
$^{9}$Department of Physics and Astronomy, Amherst College, C025 New Science Center, 25 East Dr., Amherst, MA 01002-5000, USA\\
$^{10}$Department of Physics and Astronomy, Vanderbilt University, Nashville, TN 37235, USA\\
$^{11}$NASA Goddard Space Flight Center, Exoplanets and Stellar Astrophysics Laboratory (Code 667), Greenbelt, MD 20771, USA}

\pubyear{2023}

\begin{document}
\label{firstpage}
\pagerange{\pageref{firstpage}--\pageref{lastpage}}
\maketitle

\begin{abstract}
We analyze high-cadence data from the Transiting Exoplanet Survey Satellite (TESS) of the ambiguous nuclear transient (ANT) ASASSN-18el. The optical changing-look phenomenon in ASASSN-18el has been argued to be due to either a drastic change in the accretion rate of the existing active galactic nucleus (AGN) or the result of a tidal disruption event (TDE). Throughout the TESS observations, short-timescale stochastic variability is seen, consistent with an AGN. We are able to fit the TESS light curve with a damped-random-walk (DRW) model and recover a rest-frame variability amplitude of $\hat{\sigma} = 0.93 \pm 0.02$ mJy and a rest-frame timescale of $\tau_{DRW} = 20^{+15}_{-6}$ days. We find that the estimated $\tau_{DRW}$ for ASASSN-18el is broadly consistent with an apparent relationship between the DRW timescale and central supermassive black hole mass. The large-amplitude stochastic variability of ASASSN-18el, particularly during late stages of the flare, suggests that the origin of this ANT is likely due to extreme AGN activity rather than a TDE.
\end{abstract}

\begin{keywords}
accretion, accretion discs --- black hole physics --- galaxies: active --- galaxies: nuclei
\end{keywords}

\section{Introduction}

Active galactic nuclei (AGNs) are the actively accreting supermassive black holes (SMBHs) found at the centers of $\sim$$1-5$\% of galaxies in the local universe \citep[e.g.,][]{ho08, haggard10, lacerda20}. For decades, it has been known that AGNs vary stochastically both photometrically \citep[e.g.,][]{ulrich97, drake09, macleod12} and spectroscopically \citep[e.g.,][]{bianchi05}. Due to the recent growth of optical transient surveys, AGNs and quiescent galaxies alike have also been found to undergo extreme flaring behavior. Such events include tidal disruption events \citep[TDEs;][]{vanvelzen11, gezari12b, holoien14a, holoien16a, payne21, hinkle21a}, dramatic changes in the accretion rates of known AGNs \citep[e.g.,][]{denney14, shappee14, wyrzykowski17, trakhtenbrot19a, frederick19, frederick21}, and a growing class of ambiguous nuclear transients \citep[ANTs;][]{neustadt20, hinkle21c, holoien21}.

Broadly, nuclear transients share a set of observed characteristics including bright ultraviolet (UV) emission \citep[e.g.,][]{holoien16a, neustadt20, hinkle21c}, strong emission lines \citep{holoien14a, trakhtenbrot19a, leloudas19}, slow and often smooth photometric evolution \citep{frederick21, hinkle20a, vanvelzen21}, and, for some events, X-ray emission \citep{holoien16a, neustadt20, hinkle21a}. The different classes of nuclear transients often have distinct properties in their X-ray continua, color evolution, and the presence of light curve rebrightening episodes \citep[e.g.,][]{frederick21}. However, in the case of ANTs, the lack of expected features or the presence of unexpected features can make a definitive classification difficult \citep[][]{neustadt20, frederick21, hinkle21c, malyali21, holoien21}.

One example of an ANT is ASASSN-18el \citep[AT2018zf\footnote{\url{https://www.wis-tns.org/object/2018zf}};][]{nicholls18, trakhtenbrot19b}. ASASSN-18el, $(\alpha,\delta) =$ (19:27:19.630, $+$65:33:53.78), was discovered by the All-Sky Automated Survey for Supernovae \citep[ASAS-SN;][]{shappee14, kochanek17} on 2018 March 3 in a known AGN, 1ES 1927+654 \citep{giacconi79}. Archival spectra were that of a ``true'' Type 2 AGN, with narrow lines and no broad lines even in polarized light \citep{boller03, tran11}. Pre-discovery imaging from the Asteroid Terrestrial-impact Last Alert System \citep[ATLAS;][]{tonry18} indicates that the source became active as early as 2017 December 23 \citep{trakhtenbrot19b}.

Spectroscopic follow-up of the flare found that the AGN had broad emission lines, suggesting that ASASSN-18el was an optical changing-look AGN \citep[CL-AGN, e.g.,][]{shappee14, denney14, lamassa15}, changing from a Seyfert 2 to a Seyfert 1 over the course of several months. During the optical flare, the X-ray flux dropped by several orders of magnitude \citep{trakhtenbrot19b, ricci21, laha22} and the dominant X-ray component softened from a hard power-law typical of an AGN \citep[e.g.,][]{ricci17} to a more TDE-like blackbody spectrum \citep[e.g.,][]{auchettl17}. Based on the X-ray evolution and smooth UV/optical flare, \citet{ricci20} speculated that ASASSN-18el was a TDE. Nonetheless, the overall properties of ASASSN-18el make a definitive distinction between an AGN flare and TDE difficult.

In this paper, we analyze photometry of ASASSN-18el from the Transiting Exoplanet Survey Satellite \citep[TESS;][]{ricker15}, additional late-time optical light curves, and late-time X-ray light curves from the Neutron Star Interior and Composition Explorer \citep[\textit{NICER};][]{gendreau12} to understand the emission of ASASSN-18el in the growing context of nuclear flares. This paper is organized as follows. In Section \ref{data} we detail the data used in this work. In Section \ref{analysis} we present our analysis and then discuss our results in Section \ref{disc}. Finally, we summarize our findings in Section \ref{summary}. Throughout the paper, we assume a cosmology of $H_0$ = 69.6 km s$^{-1}$ Mpc$^{-1}$, $\Omega_{M} = 0.29$, and $\Omega_{\Lambda} = 0.71$ \citep{wright06, bennett14}. The source redshift of z = 0.01905 implies a luminosity distance of 83.3 Mpc and the Galactic foreground extinction in the direction of ASASSN-18el is $A_V = 0.236$ mag \citep{schlafly11}. \citet{trakhtenbrot19b} measured a virial mass of $M_{BH} = 1.9 \times 10^{7}$ \msun\ using spectra from their follow-up campaign, which we adopt in this manuscript. As this is a single epoch measurement, we will assume an uncertainty of 0.3 dex \citep[e.g.,][]{guo20}.

\section{Data}\label{data}

Data near peak emission and up to roughly 400 days after were presented originally in \citet{trakhtenbrot19b}, with additional X-ray data shown in \citet{ricci21} and X-ray/UV data shown in \citet{laha22}. Here we re-reduce ground-based survey data on ASASSN-18el from ASAS-SN and ATLAS, now extending roughly 1300 days after peak. We additionally use TESS, Swift, NICER, and ground-based follow-up data. In this section we describe the procedures used to obtain and reduce these data.

\begin{figure*}
\centering
 \includegraphics[width=1.0\textwidth]{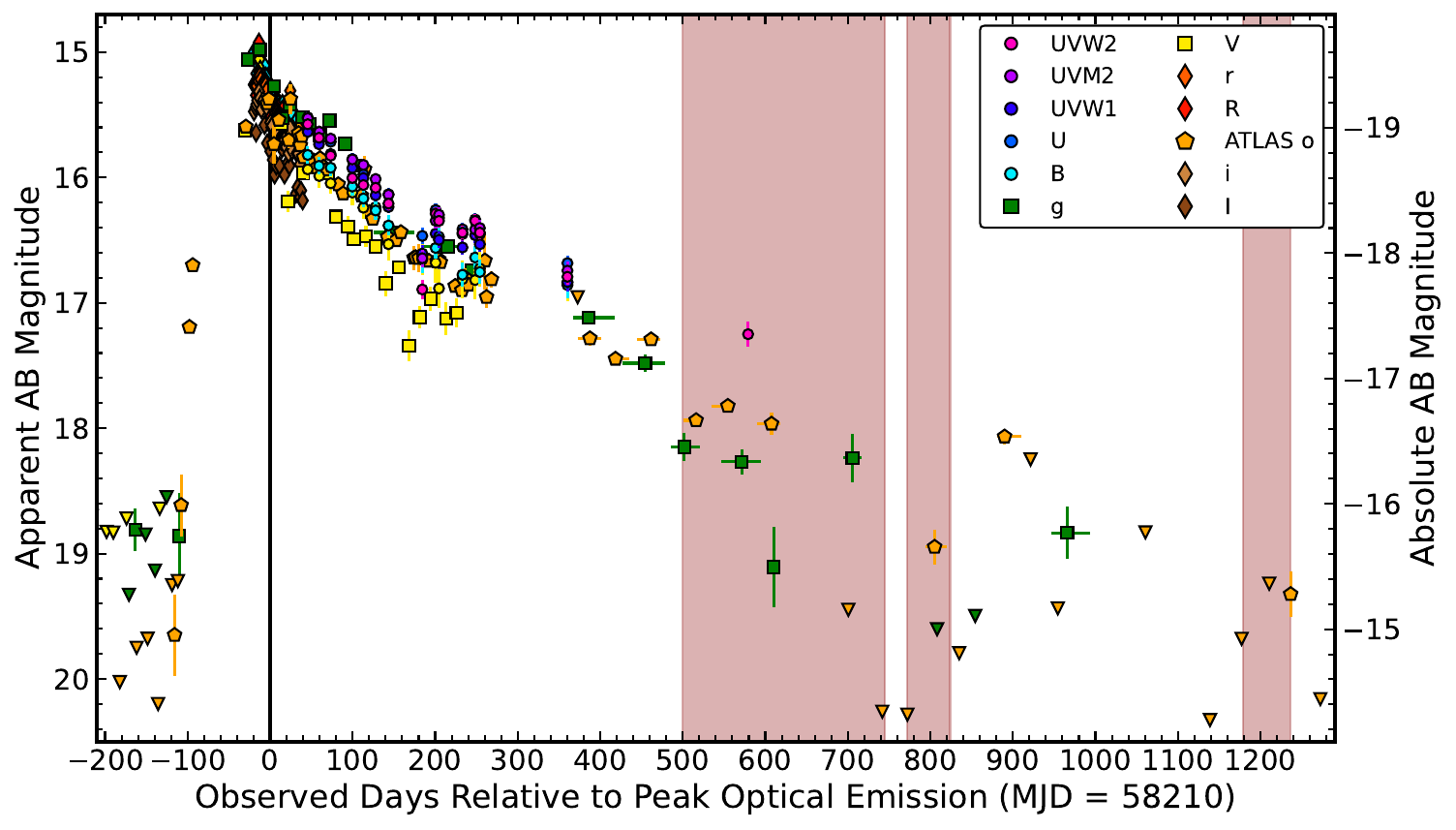}\hfill
 \caption{Optical and UV light curves of ASASSN-18el, showing ASAS-SN ($gV$; squares), ATLAS ($o$; pentagons), Swift ($UBV + UV$; circles), and other ground-based ($BVrRiI$; diamonds) photometry. The photometry shown here spans from roughly 200 days prior to peak optical emission in the ATLAS $o$-band to roughly 1250 days after in observer-frame days. The vertical red bands indicate time periods with TESS coverage. The vertical black line indicates the time of the ATLAS $o$-band peak. Horizontal error bars indicate the date range of observations stacked to obtain deeper limits and higher S/N detections, although for some data they are smaller than the symbols. Downward-facing triangles indicate 3$\sigma$ upper limits. All data are corrected for Galactic extinction and are shown in the AB system.}
 \label{fig:optical_lc}
\end{figure*}

\subsection{ASAS-SN Data}

The ASAS-SN transient survey consists of five sites located at Haleakal\=a Observatory, McDonald Observatory, the South African Astrophysical Observatory (SAAO), and two at Cerro Tololo Inter-American Observatory (CTIO), all hosted by the Las Cumbres Observatory \citep{brown13}. Each site hosts four telescopes on a common mount, and the telescopes are 14-cm aperture Nikon telephoto lenses with 8\farcs{0} pixels. However, at the time of the discovery of ASASSN-18el, ASAS-SN operated with two units taking $V$-band images and three units taking $g$-band images. After the seasonal break following the transient peak emission, ASAS-SN had fully switched to $g$-band imaging and all of the later time ASAS-SN data are in the $g$-band.

The images were reduced using the automated ASAS-SN pipeline \citep{shappee14}, which uses the ISIS image subtraction package \citep{Alard1998, Alard2000}. We then used the IRAF {\tt apphot} package with a 2-pixel radius (approximately 16\farcs{0}) aperture to perform aperture photometry on each subtracted image, to produce a differential light curve. We calibrated our photometry using the AAVSO Photometric All-Sky Survey \citep{Henden2015}. To obtain good detections or limits throughout the transient evolution, we stacked the ASAS-SN epochs. We used 10-day bins prior to and during the flare for both $V$- and $g$-band data, and 50-day bins for the $g$-band data after the seasonal break to sample the decline. We excluded ASAS-SN data more than 1000 days after peak optical emission as contamination from the residuals of a nearby star became similar in amplitude to the transient.

\subsection{ATLAS Data}

At the time of discovery, ATLAS consisted of two 0.5-m f/2 Wright Schmidt telescopes located on Haleakal\=a and at the Mauna Loa Observatory. The ATLAS telescopes typically obtain four 30-second exposures of 200--250 fields, covering roughly a quarter of the sky visible from Hawai'i, each night \citep{smith20}. Depending on the Moon phase, ATLAS uses either the `cyan' ($c$) filter from 420--650 nm or the `orange' ($o$) filter covering the 560--820 nm range \citep{tonry18}. As there was little $c$-band coverage of the flare, particularly in the late phases where we focus our study, we have excluded it from our analysis.

The ATLAS images are processed by a pipeline that includes flat-field, astrometric, and photometric calibrations. We performed forced photometry on the subtracted ATLAS images of ASASSN-18el as described in \citet{tonry18} to create a differential light curve. Due to the bright host galaxy and nearby stars, there are many epochs with artifacts remaining in the subtracted images, as indicated by the high reduced $\chi^2$ of the PSF fits. We therefore excluded images with reduced $\chi^2 > 10$. Similar to our ASAS-SN light curves, we combined individual epochs to increase the S/N. Prior to the flare, we combined the data in 10-day bins, during the flare we combined the four intra-night observations, and after the seasonal break we stacked in 30-day bins.

\subsection{Additional Photometry}

We also obtained $BVrRiI$ data from several ground-based observatories. These include the Las Cumbres Observatory Global Telescope (LCOGT) network \citep{brown13}, the 0.5-m DEdicated MONitor of EXotransits and Transients \citep[DEMONEXT;][]{villanueva16, villanueva18}, the 0.5-m Iowa Robotic Telescope, and the Post Observatory.

We reduced these data using standard procedures including flat-field corrections and then obtained the astrometry for each image using astrometry.net \citep{barron08,lang10}. For these ground-based data we used {\tt apphot} to measure 5\farcs{0} aperture magnitudes of the host plus transient emission, and subtracted the 5\farcs{0} host flux computed by \citet{hinkle21b}. We used \textsc{refcat} \citep{tonry18b} magnitudes of stars in the field and the corrections of \citet{lupton05} for calibration. We stacked all of this ground-based photometry in 1 day bins as the source was bright and there were several facilities taking images concurrently. 

We also use the Swift $UBV + UV$ photometry of ASASSN-18el presented in \citet{hinkle21b}. Some of these data were originally presented in \citet{trakhtenbrot19b}, but were re-reduced in \citet{hinkle21b} to account for corrected calibration files for the Swift UV filters. Figure \ref{fig:optical_lc} shows the light curves of ASASSN-18el from ASAS-SN, ATLAS, Swift and ground-based observatories in AB magnitudes and corrected for Galactic foreground extinction.

\subsection{\textit{TESS} Observations}

The host galaxy of ASASSN-18el lies close to the North Ecliptic Pole and the TESS northern continuous viewing zone. It was observed in each sector of Cycle 2 except for Sector 24. The source was very close to the edge of the chip in Sector 14, and therefore we do not include this sector. This allowed us to get a high-cadence and high-precision light curve beginning roughly 500 days after the peak optical emission. The source was also observed again in Sectors 40 and 41 of Cycle 4.

The differential light curves are generated from the TESS full frame images (FFIs) following the procedures of \citet[][]{vallely19, vallely21}. Because the TESS PSF has non-trival structure and the camera orientations rotate between sectors, we constructed independent reference images for each sector. We selected the first 100 FFIs of good quality obtained in each sector, excluding images with above average sky background levels or PSF widths. We also excluded FFIs that were affected by an instrument anomaly, showed significant backgrounds due to scattered light, had data quality flags, or a compromised spacecraft pointing. We converted the measured fluxes into TESS-band magnitudes using an instrumental zero point of 20.44 electrons per second in the FFIs \citep[TESS Instrument Handbook; ][]{vanderspek18}. 

\subsection{\textit{NICER} Observations}

Throughout the evolution of ASASSN-18el, X-ray observations were obtained using NICER \citep{gendreau12}, an external payload on the International Space Station. NICER observed ASASSN-18el a total of 431 times between 2018 May 22.8 and 2021 February 17.7 (ObsIDs: 1200190101$-$1200190287, 2200190201$-$2200190370, and 3200190301$-$3200190451; PI: Kara).

The NICER data were reduced using \textsc{NICERDAS} version 6a and \textsc{HEASOFT} version 6.26.1. We applied standard filtering criteria in \textsc{nicerl2}. These criteria include\footnote{See \url{https://heasarc.gsfc.nasa.gov/docs/nicer/data_analysis/nicer_analysis_guide.html} or \citep{2019ApJ...887L..25B} for more details.}: a \textit{NICER} pointing of ANG\_DIST $<$0.015 degrees from the source location and exclusion of events during a South Atlantic Anomaly passage or when Earth was 30$^{\circ}$ (40$^{\circ}$) above the dark (bright) limb. We also removed events which were flagged as overshoot or undershoot events (EVENT\_FLAGS=bxxxx00), and used the ``trumpet filter'' to remove events with a PI\_RATIO $>$ 1.1+120/PI \citep{2019ApJ...887L..25B}. 

To extract count rates for each epoch, we used \textsc{XSELECT}, and the standard ancillary response (ARF; nixtiaveonaxis20180601v002.arf) and response matrix (RMF; nixtiref20170601v001.rmf) files from the \textit{NICER} CALDB. We assumed a photon index of 1.7 and the Galactic column density of $6.4 \times 10^{20}$ cm$^{-2}$ \citep{ricci20} to convert the count rates to fluxes. Data within $\sim$500 days of peak optical emission were presented in \citet{ricci21}. Here we extend the NICER data coverage until nearly 1100 days after peak to trace the X-ray luminosity evolution at late times.

\section{Analysis}\label{analysis}

\subsection{Optical Light Curve}

In addition to the pre-flare and near-peak data presented in \citet{trakhtenbrot19b}, the light curves shown in Figure \ref{fig:optical_lc} now extend several years after the peak emission of the transient. It is clear that at the beginning of the TESS observations of ASASSN-18el, the transient is still detectable in both the stacked ASAS-SN $g$-band and the ATLAS $o$-band data.

\begin{figure}
\centering
\includegraphics[width=0.48\textwidth]{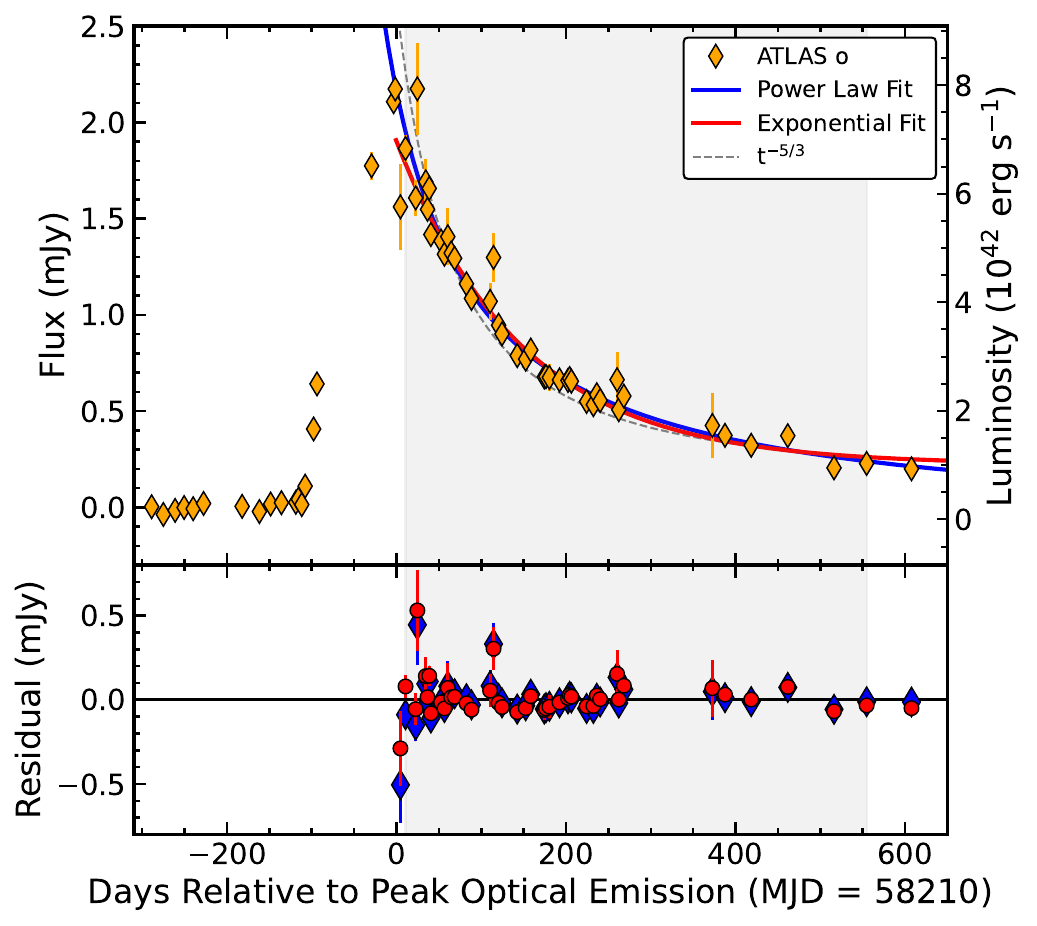}\hfill
 \caption{\textit{Top panel}: ATLAS $o$-band light curve (orange diamonds) and the best-fitting power-law (exponential-decay) model in blue (red). The light gray shaded region shows the time period included in the fits. Also shown with the dashed gray line is a $t^{-5/3}$ decline assuming the same $t_{0}$ as the free index power-law fit. \textit{Bottom panel}: The residuals of the models, shown in the same color as the fits themselves.}
 \label{fig:decline}
\end{figure}

\begin{figure*}
\centering
 \includegraphics[width=0.99\textwidth]{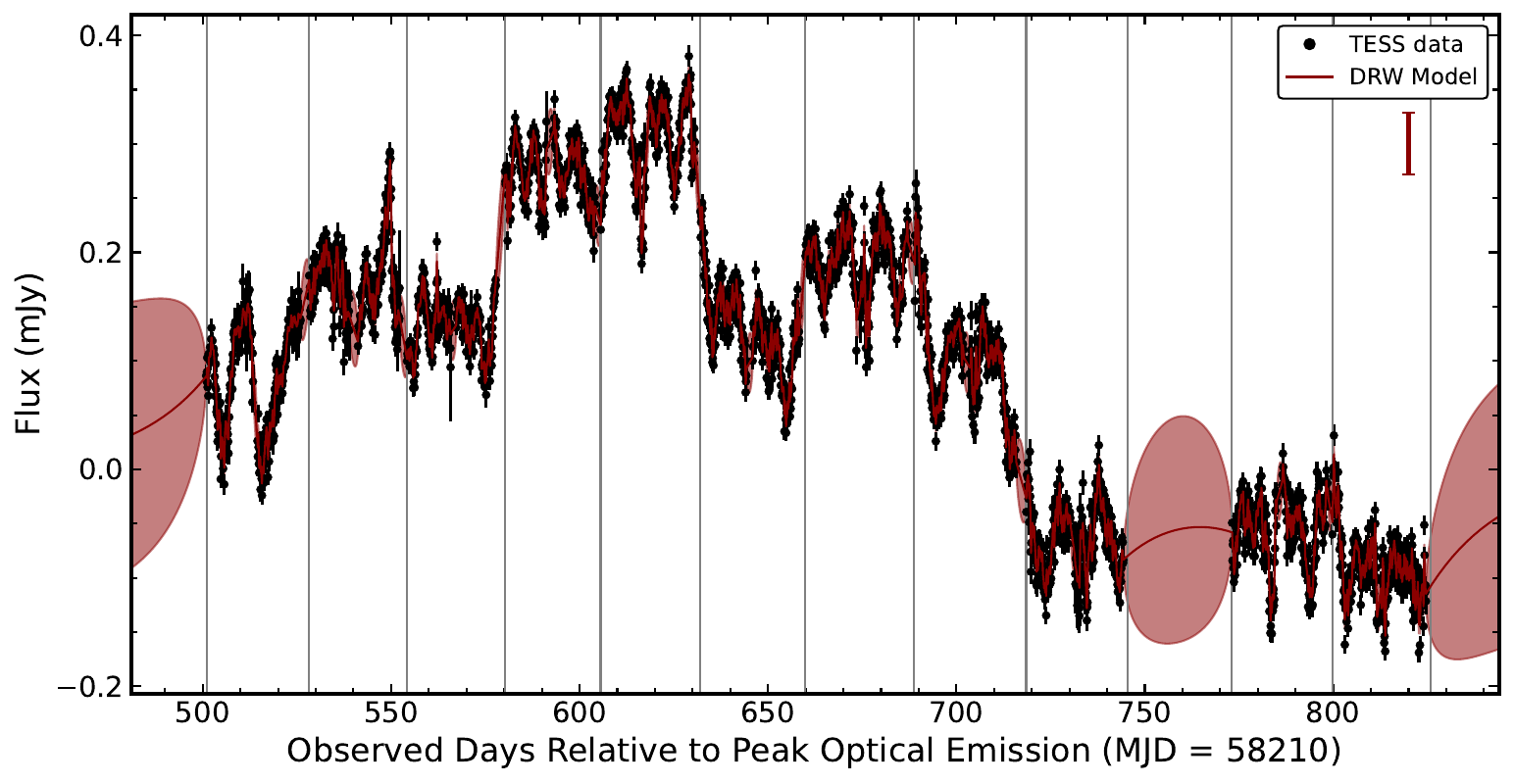}\hfill
 \caption{TESS data stacked in 2-hour bins (black points) with the mean DRW light curve shown in red. In this model, linear offsets are computed between the sectors such that the global mean flux is zero. Here, no correction for Galactic extinction has been applied. The shading represents the RMS of the DRW light curves. The range of allowed light curves broadens rapidly when there is no data in a given time interval. The vertical gray bars mark the beginning of new sectors. The inter-sector calibration has larger errors than the point-to-point precision in the TESS light curve (shown as the error bar in the upper right) but these uncertainties are included in our analysis and do not affect our results.}
 \label{fig:javelin_lc}
\end{figure*}

\begin{figure}
\centering
 \includegraphics[width=0.48\textwidth]{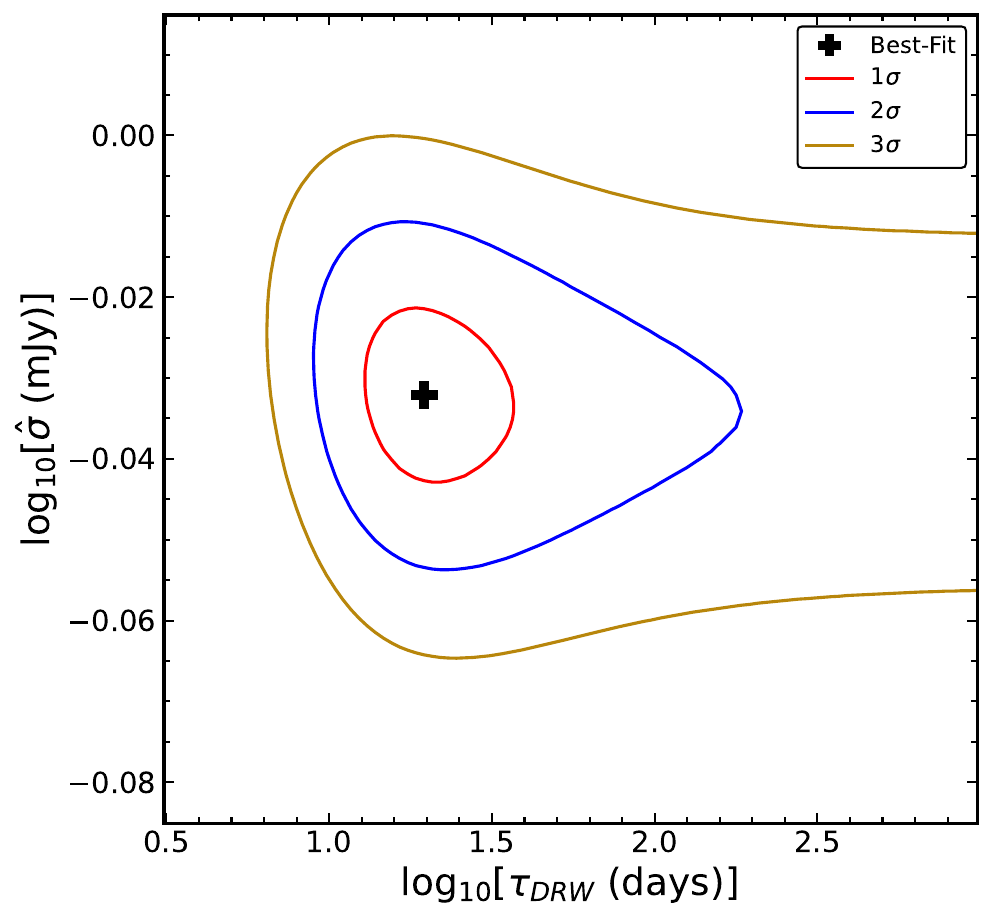}\hfill
 \caption{Contours of log($\hat{\sigma}$) and log($\tau_{DRW}$), shown at 1, 2, and 3$\sigma$ from the best-fit value shown as a black cross. The values of $\hat{\sigma}$ and $\tau$ have been corrected into the rest frame using the host redshift of z = 0.01905 and $\hat{\sigma}$ has been corrected for Galactic extinction.}
 \label{fig:sig_tau_hist}
\end{figure}

To obtain a time of peak optical light, we fit a parabola to the ATLAS $o$-band light curve as it is the most complete on both sides of the peak. Since the cadence is low, we fit the parabola over a wide date range of dates between MJD = 58116.2 and MJD = 58292.5. We generated 10,000 realizations of the ATLAS light curve over this date range with each flux perturbed by its uncertainty assuming Gaussian errors. We then fit a parabola to each of the perturbed light curves and took the median value as the peak of the distribution and used the 16th and 84th percentiles as the uncertainties. This leads to a time of peak $o$-band emission of MJD $= 58210 \pm 2$ with a peak flux of $1.9 \pm 0.1$ mJy. Several studies of TDE flares have found significant offsets between the peak times in different optical bands \citep{holoien19a, holoien20, hinkle21a} although the cadence of ASAS-SN data and the lack of UV data near peak preclude such comparisons.

One distinguishing feature of TDE light curves is that they are commonly described by a power-law decline in flux. In particular, when the mass per unit binding energy is constant, the decline slope is $t^{-5/3}$ \citep[e.g.,][]{evans89}. To examine the decline slope of ASASSN-18el in the context of TDEs, we fit the ATLAS $o$-band flux as

\begin{equation}
f = f_{10} \bigg(\frac{t - t_{0}}{10 \text{ days}}\bigg)^\alpha + h
\end{equation}

\noindent to estimate the time $t_{0}$, the flux 10 days later $f_{10}$, the flux offset $h$, and the power-law index $\alpha$. We fit the ATLAS data from 10 days after peak to MJD = 58817.5. An MCMC fit yields best-fit parameters of $t_{0} \text{(MJD)} = 58101^{+17}_{-14},\ f_{10} = 28^{+12}_{-10} \  \text{mJy},\ h = -0.10^{+0.06}_{-0.09} \  \text{mJy},\ \text{and} \ \alpha = -1.05^{+0.15}_{-0.12}$ and this model is shown in Figure \ref{fig:decline}. This decline is consistent with the UV decline slope of $\alpha = -0.91 \pm 0.04$ measured by \citet{laha22}. The decline slope is also significantly flatter than expected for TDEs \citep[e.g.,][]{rees88, phinney89}, consistent with the qualitative comparison shown in \citet{trakhtenbrot19b}. In Fig. \ref{fig:decline} we also show a $t^{-5/3}$ decline, fit with the same $t_{0}$ as above. While the $t^{-5/3}$ decline roughly follows the ATLAS photometry, the flatter decline slope of $\alpha = -1.05$ clearly provides a better description of the decline seen for ASASSN-18el.

The root mean squared (RMS) residuals from the power-law fit are 0.11 mJy after subtracting the median ATLAS flux uncertainty \citep[e.g.,][]{vaughan03}. For the remaining calculations of the RMS scatter, we subtract the median flux uncertainty to correct for the contribution of statistical noise to the observed variability.

We next fit the decline of ASASSN-18el using an exponential-decay model

\begin{equation}
f = a e^{-(t - t_{peak}) / \tau_{decay}} + c
\end{equation}

\noindent similar to what has been done for some TDEs \citep[e.g.,][]{holoien14b, holoien16a, holoien16b, payne21}. Here we set $t_{peak}$ of MJD = 58210 from our fit to the ATLAS $o$-band light curve and fit the ATLAS data over the same time period to find best-fit parameters of $a = 1.69^{+0.07}_{-0.06} \  \text{mJy}, \ \tau_{decay} = 147^{+11}_{-10} \ \text{days}, \ \text{and} \ c = 0.22 \pm 0.02 \ \text{mJy}$. This model is shown in Figure \ref{fig:decline} and it also provides a reasonable description of the data, with RMS residuals of 0.15 mJy. The best-fit $\tau_{decay} \sim 147$ days is roughly three times longer than found for typical TDEs \citep{holoien14b, holoien16a, holoien16b}. The exponential decay fit is also a better description of the data than a  $t^{-5/3}$ decline.

We next compare the RMS measurements obtained from the decline fits to the RMS scatter in the archival ATLAS light curve. Only considering data taken before the beginning of the rise on MJD $\simeq$ 58110, we find an RMS scatter of 0.02 mJy in the ATLAS o-band data, suggesting that the transient is significantly more variable than the host galaxy was in quiescence.

\subsection{TESS Light Curve}

\begin{figure*}
\centering
\includegraphics[width=.48\textwidth]{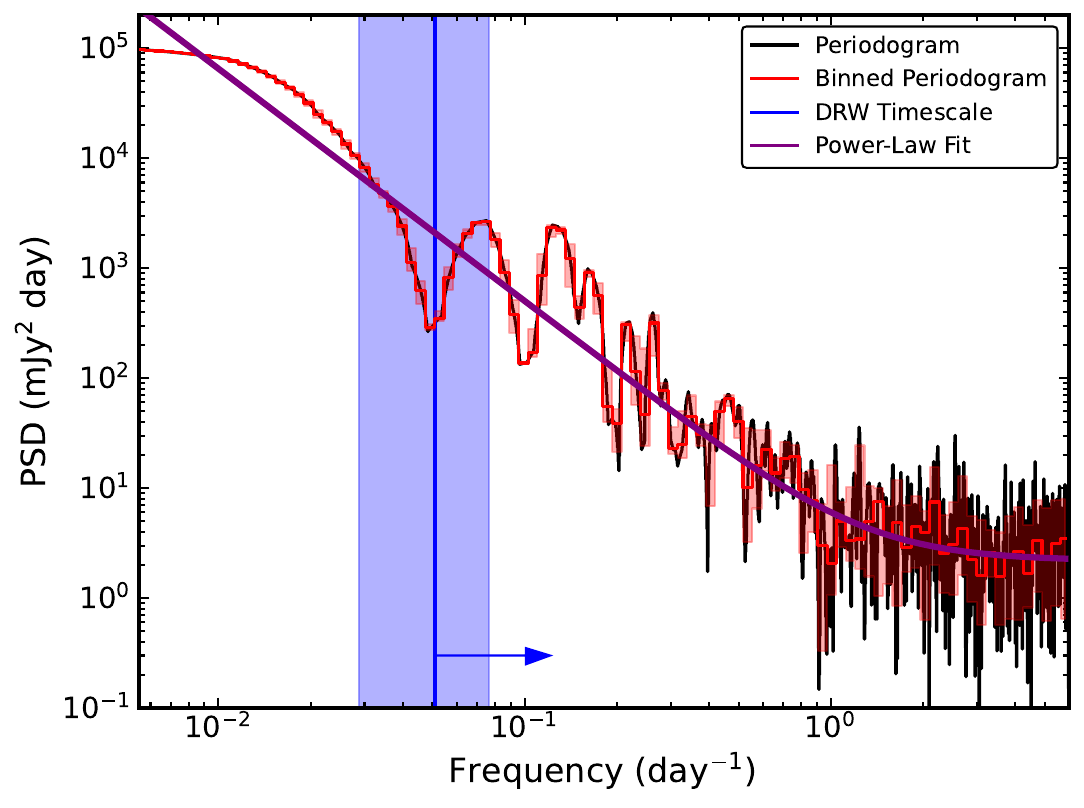}\hfill
\includegraphics[width=.48\textwidth]{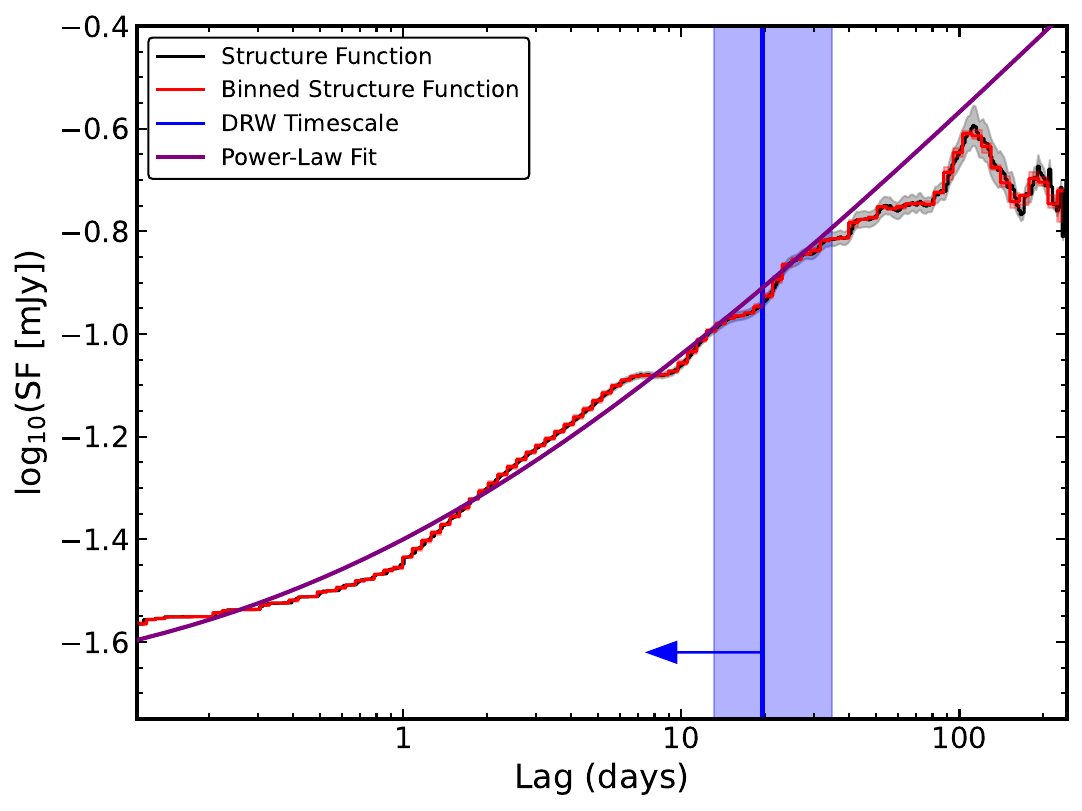}\hfill
 \caption{Power spectral density (left panel) and structure function (right panel) of the 2-hour binned TESS light curve. The PSD is computed using \textsc{scipy.signal.periodogram}. The red lines are the PSD and SF binned in 100 logarithmic bins and the red shading indicates the 1$\sigma$ scatter over that bin width. The gray band around the SF is the uncertainty at that lag. The blue lines are the estimated $\tau_{DRW}$, with the shading representing the uncertainty. As expected, the best-fit $\tau_{DRW}$ roughly coincides with the break in the PSD from a flat white noise to the higher frequency power-law slope. The purple lines are a power-law fits, with the PSD fit giving $\alpha = -2.11$ and the SF fit yielding $\gamma = 0.54$, both consistent with the expectations of a DRW model. The blue arrows represent the regions over which the power-law fits are done.}
 \label{fig:psd}
\end{figure*}

With our TESS light curve, we are able to study the variability properties of ASASSN-18el in great detail. In order to understand this variability in the context of AGN, we used a damped random walk (DRW) model \citep[e.g.,][]{kelly09, kozlowski10, macleod10, zu11, macleod12, zu13} to fit the variability seen in Figure \ref{fig:javelin_lc}. Fundamentally, a DRW variability model assumes that the power spectral density (PSD) of an AGN can be described by a flat spectrum (white noise) at low frequencies, with a smooth transition to a $f^{-2}$ power-law (red noise) at higher frequencies. The parameters of the model are the characteristic timescale $\tau_{DRW}$ and the amplitude $\hat{\sigma} = \sigma[(2 \times 365.25 \textrm{ d}) / 
 \tau_{DRW}]^{1/2}$. 

Following \citet{kozlowski10}, we fit the observed TESS light curve from Sectors 15-26 allowing for an independent flux offset for each sector (these are simply additional linear parameters in the model). This provides a well-defined statistical means of including the non-adjacent data in Sectors 25-26, drives the mean flux to zero, and includes all the uncertainties from these flux offsets into the estimates of $\hat{\sigma}$ and $\tau_{DRW}$. The magnitude of the median offset is 57 $\mu$Jy with a dispersion of 83 $\mu$Jy. We binned our TESS light curve in 2-hour bins so that the light curve has a reasonable signal-to-noise ratio per binned epoch. Throughout the remainder of this manuscript, we use these flux offsets to calibrate the fluxes for each individual Cycle 2 sector relative to the first TESS sector. However, we note that the inter-sector calibration likely has larger errors than the point-to-point precision in the TESS light curve, and long timescale trends should be viewed with caution.

The error contours for the variability timescale and amplitude are shown in Figure \ref{fig:sig_tau_hist}, with the parameters corrected for source redshift and Galactic extinction. At the 1$\sigma$ level, the contours are well-behaved, but the allowed upper bound on $\tau_{DRW}$ increases rapidly at higher confidence. The characteristic timescale of $\tau_{DRW} = 20^{+15}_{-6}$ days is similar to those found for other AGNs with similar mass SMBHs \citep[e.g.,][]{kelly09, burke21}. The variability amplitude of $\hat{\sigma} = 0.93 \pm 0.02$ mJy, corresponding to $\sigma \sim 0.13$ mJy for $\tau_{DRW} = 20$ days, is large, immediately suggestive of AGN-like variability \citep[e.g.,][]{vandenberk04, macleod10, kozlowski16}.

We tested how our decisions on connecting individual TESS sectors together affected our results, finding that regardless of how we match adjoining sectors, the recovered $\tau_{DRW}$ values are consistent within $1\sigma$ uncertainties. The variability amplitude $\sigma$ is much more sensitive to choices on sector matching, but we are able to directly compute the RMS scatter for each TESS sector, circumventing the DRW modeling and sector-matching entirely. The median RMS scatter per sector is 24 $\mu$Jy, which is 8\% of the median TESS flux over the same range.

Figure \ref{fig:psd} shows the PSD and structure function (SF) of the TESS light curve of ASASSN-18el between Sectors 15 and 23, avoiding the sector-long gaps in the TESS coverage. The PSD was computed using \textsc{scipy.signal.periodogram} and a flat top window function to increase amplitude accuracy. As expected from previous studies of AGN variability \citep[e.g.,][]{mushotzky11, kozlowski17}, the PSD is flat at low frequencies, with a declining power-law at higher frequencies. Similarly, the SF is roughly flat at long lags with a declining slope towards shorter lags. At the highest frequencies (shortest lags), the PSD (SF) flattens again as the signal becomes dominated by photometric noise. There is also a dip in the PSD at $\sim 0.04$ day$^{-1}$, likely due to the 27 day length of the TESS sector and the linear offsets applied between the sectors. If we fit the PSD above the measured value of $\tau_{DRW}$ (in frequency) with a single power-law plus a photometric noise term, we find a power-law index of $\alpha = -2.11 \pm 0.16$.  Likewise, if we fit the SF below the measured value of $\tau_{DRW}$ (in time) with the same model we find a power-law index of $\gamma = 0.54 \pm 0.02$.

The canonical power-law slope for a DRW model is $\alpha = -2$ \citep[e.g.,][]{zu11, zu13} for the PSD and $\gamma = 0.5$ \citep[e.g.,][]{kozlowski16} for the SF. Studies of AGNs using high-quality Kepler light curves \citep{mushotzky11, smith18} have shown a range of power-law slopes between $\alpha = -1.7$ and $\alpha = -3.3$, although \citet{moreno21} showed that Kepler systematics can mimic AGN variability. Therefore, we find that the power-law slopes for the PSD and SF of ASASSN-18el are fully consistent with other optical AGNs and with the theoretical assumptions made when fitting the TESS light curve with a DRW model.

\subsection{Optical/X-ray Correlation}

\begin{figure}
\centering
\includegraphics[width=0.48\textwidth]{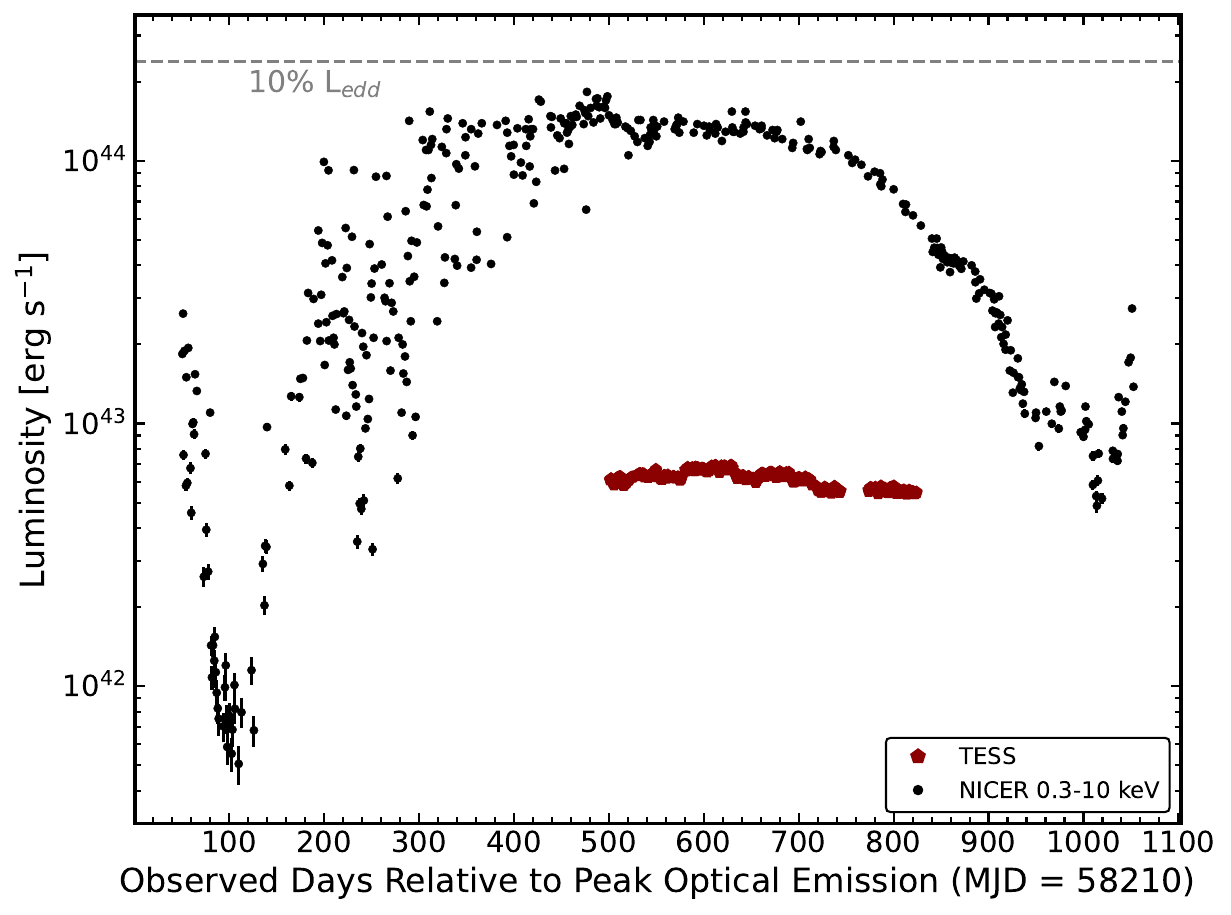}\hfill
 \caption{NICER (black circles) and TESS (dark red pentagons) light curves of ASASSN-18el. The NICER luminosity is integrated over the 0.3-10 keV bandpass. The TESS luminosity is $\lambda$L$_{\lambda}$ at the pivot wavelength of the TESS bandpass and includes the quiescent host galaxy plus AGN flux. It is clear that the X-ray component of the emission is more than an order of magnitude more luminous than the optical component during the time of overlapping data.}
 \label{fig:nicer}
\end{figure}

Next we searched for a correlation between the TESS optical and NICER X-ray light curves, shown in Figure \ref{fig:nicer}. Here we include the quiescent host galaxy plus AGN flux in the TESS bandpass to the zero-mean TESS light curve shown in Fig. \ref{fig:javelin_lc}. The X-ray emission is roughly an order of magnitude higher than the optical emission at the same phase, unlike many other ANTs \citep{neustadt20, hinkle21c}. Qualitatively, both light curves show a plateau over the time period from 500 to 700 days after the peak emission where they overlap.

We tried to measure a temporal lag between the NICER and TESS light curves. If the optical variability was a result of the reprocessing of the far more luminous X-rays, we would expect to see a correlation between light curve features in the X-ray and the optical, with the optical lagging behind the X-ray. Assuming this is due to light travel time, this lag would give an approximate physical size of the accretion disk between the X-ray emitting corona and the cooler portions of the disk from which the emission seen in the TESS bandpass originate. Unfortunately, neither the straightforward cross-correlation function code \textsc{PyCCF} \citep{sun18} nor the more sophisticated DRW modeling code \textsc{javelin} \citep{zu11, zu13} model produced a useful result. Additionally, we tested if binning the TESS data and/or smoothing the light curves to remove long-term trends affected the lag fits, but neither was able to return an acceptable result.

\subsection{Very Late-time Data}

\begin{figure*}
\centering
 \includegraphics[width=1.0\textwidth]{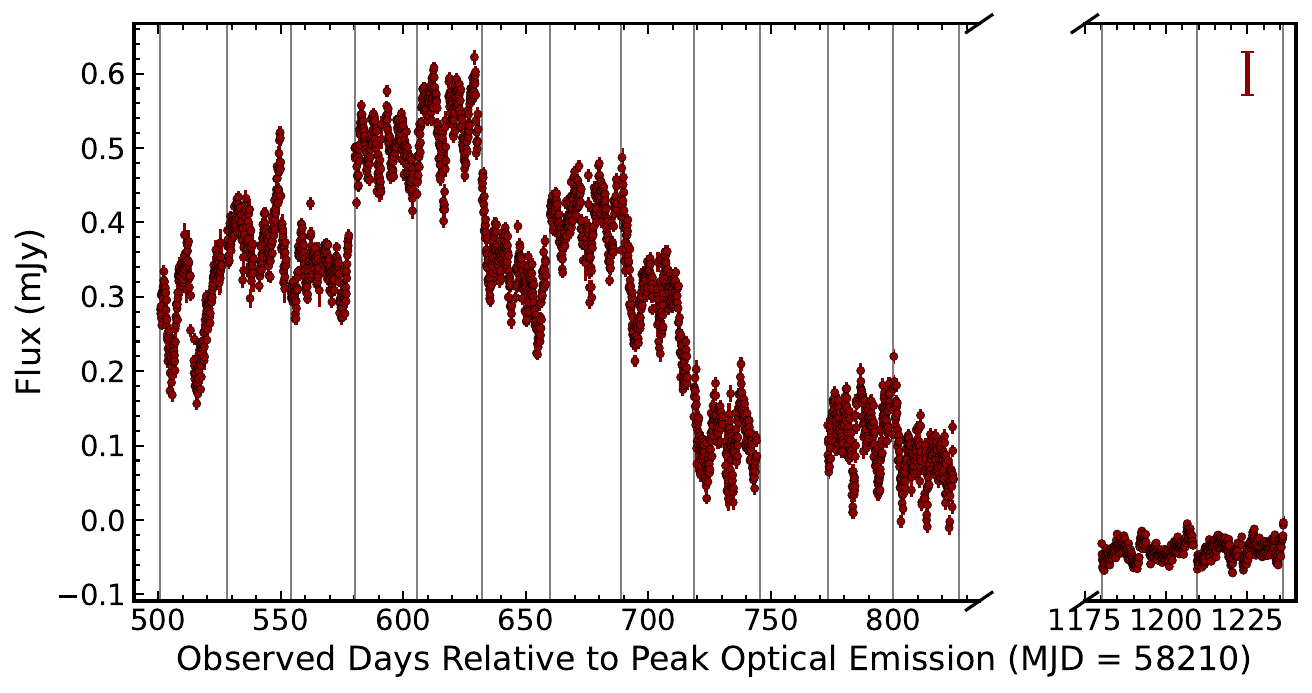}\hfill
 \caption{TESS data stacked in 2 hr bins (red points) spanning from the initial Sector 15 light curve to the very late-time Sector 41 light curve. The vertical gray bars mark the beginning of new sectors and the error bar in the upper right indicates the median magnitude of the sector-to-sector offsets. Shown for TESS sectors prior to Sector 40 is the zero-mean DRW light curve from Fig. \ref{fig:javelin_lc} scaled such that the first TESS sector matches the differential flux in concurrent ATLAS o-band data. For the very late-time sectors, the TESS light curve is scaled to match the differential flux in concurrent ATLAS o-band data.}
 \label{fig:all_tess}
\end{figure*}

After TESS returned to the Northern hemisphere in Cycle 4, ASASSN-18el was observed again. We show the full TESS light curve as of the end of 2021, including Sectors 15-23, 25-26, and 40-41 in Figure \ref{fig:all_tess}. It is immediately clear that the rapid high-amplitude variability has faded by the very-late time sectors. To quantify the change in variability amplitude between the early and late sectors, we measured the RMS scatter in the first sector and compared it to the last sector. For Sector 15, the RMS scatter around the median flux is 46 $\mu$Jy. Roughly 2 years later in Sector 41, the RMS scatter had dropped by a factor of almost 5 to 9 $\mu$Jy. ASASSN-18el will be observed in 13 continuous TESS sectors in Cycles 4 and 5 (Sectors 47 - 59), allowing for additional late-time monitoring of the source variability well after the transient has faded.

While the overall light curve shape is not particularly important for understanding the short-term variability expected of AGNs, we used stacked late-time ATLAS data to measure the $o$-band flux concurrent with active TESS sectors to allow us to create a long-term TESS light curve. We matched the first TESS sector to the concurrent ATLAS $o$-band flux and applied the linear offsets between sectors computed in the DRW modeling to calibrate the flux of the remaining Cycle 2 TESS sectors. For the Cycle 4 TESS light curves, we added the concurrent ATLAS $o$-band flux to the differential TESS light curve. As our primary aim in Figure \ref{fig:all_tess} is to show the consistent strong short-term variability of ASASSN-18el over a 2 year time span, details on sector to sector matching are largely inconsequential.

\section{Discussion}\label{disc}

Two plausible explanations have been put forward to explain the changing-look phenomenon observed for the ANT ASASSN-18el. The first is related to instabilities in the existing AGN disk, either as a result of a change in the accretion flow \citep{trakhtenbrot19b} or something more exotic like a magnetic inversion in the disk \citep{scepi21, laha22}. The second is increased accretion due to a TDE, which disrupted the existing X-ray corona while powering a large UV/optical flare \citep{ricci20, masterson22, li22}. 

The variability seen in the TESS light curve is typical of known AGNs and consistent with the DRW model frequently used to describe AGN variability. Both the PSD and SF of the TESS light curve are typical of AGNs. Therefore, independent of any modeling of the TESS light curve and choices on sector matching, this short-term stochastic variability suggests an AGN origin for ASASSN-18el.

\begin{figure}
\centering
\includegraphics[width=0.48\textwidth]{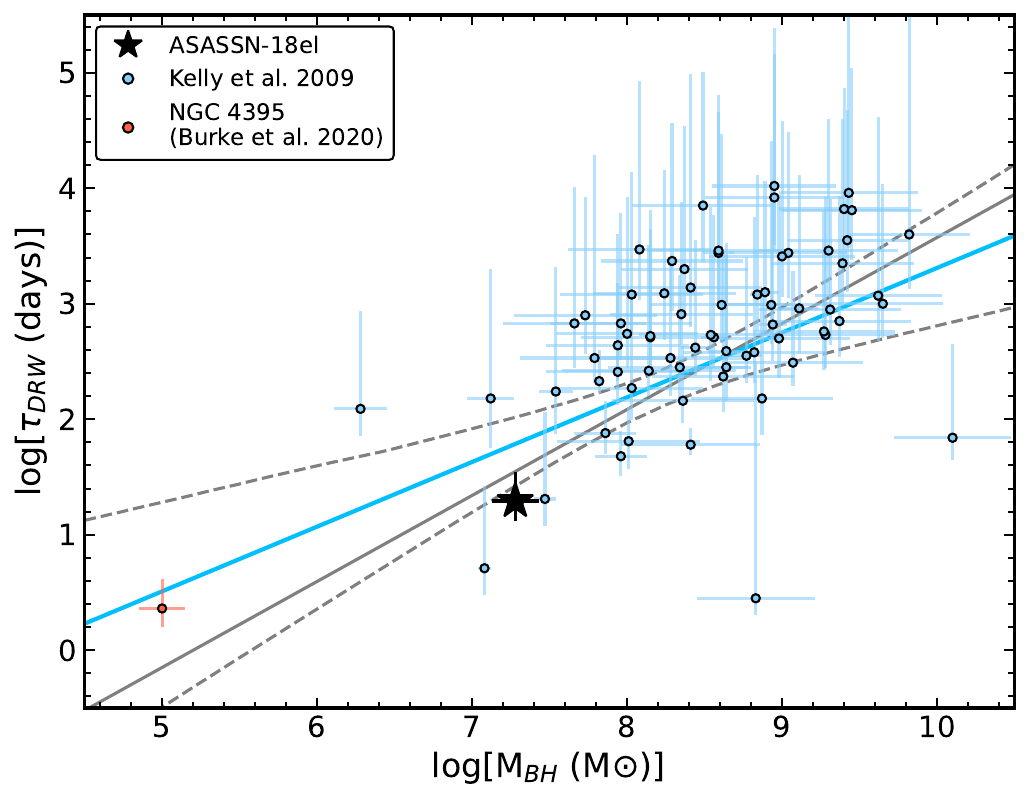}\hfill
 \caption{The rest-frame DRW timescale ($\tau_{DRW}$) as compared to the SMBH mass for a given AGN. The sample of \citet{kelly09} is shown in light blue, with their best-fit line shown with the light blue solid line. The solid gray line is our fit of the \citet{kelly09} sample using the methods of \citet{kelly07}. The dashed gray lines are plus/minus one sigma from the best-fit line. Both NGC 4395 \citep{burke20} and ASASSN-18el are excluded from this fit.}
 \label{fig:tau_mbh}
\end{figure}

Nevertheless, with a measurement of $\tau_{DRW}$, we can compare to known relationships, such as the one between $\tau_{DRW}$ and the central SMBH mass, shown in Figure \ref{fig:tau_mbh}. As the black hole mass increases, so does the characteristic timescale, resulting from the scaling of accretion disk parameters \citep[e.g.,][]{kelly09, zu13}. In particular, \citet{burke21} suggest that the characteristic variability timescale is linked to the thermal timescale for the UV-emitting region of the accretion disk, even when that $\tau_{DRW}$ is measured with optical rather than UV photometry. ASASSN-18el lies near the trend between $\tau_{DRW}$ and SMBH mass. In Figure \ref{fig:tau_mbh} we have shown both the original best-fit line and our fit to the data from \citet{kelly09} using the MCMC procedure of \citet{kelly07} to better illustrate the range of allowed slopes for this relationship given many sources with large uncertainties.

Similarly, we can compare the pre-flare constraints from ASAS-SN and ATLAS on variability to those obtained with TESS. Prior to the flare, there are only hints of weak activity, with sparse detections in the ASAS-SN $V$-band photometry. This is consistent with the archival Seyfert 2 classification of the AGN 1ES 1927+654. The lack of strong variability prior to the flare in contrast with short-timescale variability seen during the flare suggests that ASASSN-18el is an example of an increase in the accretion rate onto the AGN rather than a TDE. This scenario is also consistent with the late-time decrease in variability, as the AGN begins to return to its quiescent state.

It is important to note that there are some caveats to the analysis of AGN light curves using a DRW model. The primary concern is the length of the light curve relative to the recovered timescale. Recent simulations suggest that for robust recovery of DRW parameters, the length of the light curve should be at least ten times the $\tau_{DRW}$ \citep[e.g.,][]{kozlowski17, kozlowski21}, with shorter light curves biased towards shorter timescales. For ASASSN-18el the length of the TESS light curve is roughly 16 times longer than the recovered $\tau_{DRW}$. Figure 4 of \citet{kozlowski21} suggests that there is only a $\sim 0.1$ dex uncertainty in $\tau_{DRW}$ at this length. Even if the true $\tau_{DRW}$ for ASASSN-18el was underestimated by a few tenths of a dex, it would remain close to the trend shown in Fig. \ref{fig:tau_mbh}.  Finally, as noted in Section 3.2, results related to the long-timescale trends in the TESS light curve should be viewed with caution because of uncertainties and systematic issues associated with the inter-sector calibrations.

In addition to the high precision TESS photometry, the X-ray light curve from NICER gives important constraints on physical processes occurring close to the SMBH. Two notable aspects of the X-ray emission for ASASSN-18el are the extremely high X-ray to optical luminosity ratio and the deep dip in X-ray flux at $\sim$100 days after the optical peak. The proposed model of \citet{scepi21} may provide a natural explanation for these features. In this model, the accretion disk prior to the flare exists in a magnetically arrested disk (MAD) state \citep[e.g.,][]{narayan03}. In such a state, the UV/optical region of the disk is radiatively inefficient and the X-ray emission is synchrotron-powered \citep{scepi21}, accounting for the unusually high X-ray to optical luminosity ratio.

\citet{scepi21} suggest that a magnetic inversion occurs due to the advection of material with opposite magnetic polarity through the disk, thus disrupting the MAD state. In such a magnetic inversion, reconnection occurs first in the outer regions of the disk, increasing the radiative efficiency and powering the dramatic UV/optical flare. As the accretion flow reaches the SMBH, the X-ray corona is destroyed, decreasing the X-ray luminosity with a lag relative to the UV/optical brightening \citep{scepi21}. As magnetic flux accumulates, the X-ray corona is rebuilt and the X-ray luminosity recovers. The decline in the UV/optical emission can either be due to a newly formed MAD state or a decrease in the accretion rate. \citet{scepi21} predict that for the latter scenario, we should see a secondary decline in the X-ray luminosity roughly 2 years later. Indeed, the NICER data shows a decline in X-ray luminosity beginning at $\sim$700 days after peak optical emission, consistent with such a model.

Finally, with respect to the possibility that the changing-look phenomenon was caused by a TDE, it is difficult to explain the observed variability in the context of TDE emission. To date only a small number of TDEs have exhibited any stochastic variability \citep[e.g.,][]{holoien19a}, and TDE light curves rarely exhibit re-brightening episodes \citep[e.g.,][]{wevers23, malyali23} like the ones seen here in ATLAS and ASAS-SN photometry. Late-time studies of TDEs have suggested a shift in the emission mechanism from one tied to the fallback of mass onto the SMBH to emission from the accretion disk itself \citep{balbus18, vanvelzen19a, jonker20}. A natural assumption is that, similar to AGN disks, this may lead to enhanced variability from instabilities in disk. Nonetheless, when examining late-time data on the TDEs ASASSN-14ae and ASASSN-14li \citep{vanvelzen19a, hinkle21b}, we find that there is no significant variability on several day timescales even long after disk emission is thought to become the dominant component for these sources \citep{vanvelzen19a}. This suggests that a TDE transitioning to a disk-dominated state is unlikely to show the very large variability that we see from ASASSN-18el at similarly late times. In addition, the long overall timescale of the event is not consistent with known TDEs \citep[e.g.,][]{holoien14b, holoien16a, holoien16b, hinkle20a}. These suggest that a TDE is likely not the source of the transient emission for ASASSN-18el.

\section{Summary}\label{summary}

Through our analysis of late-time optical photometry, NICER X-ray data, and a high-cadence TESS light curve of ASASSN-18el we find the following:

\begin{itemize}

\item Optical transient emission from ASASSN-18el remains detectable above pre-flare levels for roughly 700 days after the peak optical emission with late-time re-brightening episodes continuing until at least 1250 days after peak.

\item The initial decline of ASASSN-18el as measured from the ATLAS $o$-band light curve is significantly flatter than the canonical t$^{-5/3}$ power-law for TDEs. Additionally, when the decline is fit by an exponential-decay model, the timescale of the decline is roughly three times longer than observed for typical TDEs.

\item The high-S/N and high-cadence TESS light curve within 1000 days of peak optical emission shows strong variability on timescales less than a single TESS sector, immediately suggestive of strong AGN activity. 

\item The TESS light curve's large variability amplitude and characteristic timescale are consistent with light curves of AGNs with similar SMBH masses, supporting the claim that the observed transient emission at late times is fully consistent with a dramatic AGN flare. However, there are large uncertainties and potential systematic errors related to the long-timescale trends in the TESS light curve.

\item The X-ray and optical evolution is distinct, with large dips in X-ray luminosity at early and late times despite a relatively smooth optical light curve. However, during the period of overlap between TESS and NICER data, a multi-wavelength plateau phase is observed. Nonetheless, we were unable to recover a temporal lag between the X-ray and TESS bandpasses.

\item The very late-time TESS data (more than 1000 days after peak emission) continues to show stochastic variability, but at a lower amplitude than earlier sectors.

\end{itemize}

From our analysis of these late-time data, and in particular the exquisite TESS data, we find that the flare in ASASSN-18el is likely the result of a large AGN flare, possibly induced by a magnetic inversion in the disk \citep{scepi21}. Many of the optical changing-look AGNs to date lack the wealth of data presented here. However, the few that do, such as NGC 2617 \citep{shappee14} and ASASSN-18el may suggest that such state changes are not always long term but rather may be the result of large flares.

The use of TESS and its unparalleled combination of photometric precision, cadence, and sky coverage is a promising avenue for studying the physical origin of ANTs. TESS observations of an ANT at any phase can give detailed insight into the variability properties of a galaxy. At the very least, a sector of TESS data should be able to uncover the presence of AGN-like variability in ANT host galaxies, well below what can be probed from the ground. This gives promise that we can begin to understand the range of nuclear behaviors and accretion processes through the disambiguation of ANTs.

\section*{Data availability}
	
The data underlying this article will be shared on reasonable request to the corresponding author.

\section*{Acknowledgements}
We thank the referee for helpful comments and suggestions that have greatly improved the quality of this manuscript. We also thank Alexa Anderson and Willem Hoogendam for useful comments on the manuscript.

We thank Las Cumbres Observatory and its staff for their continued support of ASAS-SN. ASAS-SN is funded in part by the Gordon and Betty Moore Foundation through grants GBMF5490 and GBMF10501 to the Ohio State University, and also funded in part by the Alfred P. Sloan Foundation grant G-2021-14192.  Development of ASAS-SN has been supported by NSF grant AST-0908816, the Mt. Cuba Astronomical Foundation, the Center for Cosmology  and AstroParticle Physics at the Ohio State University, the Chinese Academy of Sciences South America Center for Astronomy (CAS- SACA), the Villum Foundation, and George Skestos. 

JTH was supported by NASA grant 80NSSC21K0136. BJS and CSK are supported by NSF grant AST-1907570/AST-1908952. BJS is also supported by NSF grants AST-1920392 and AST-1911074. CSK is supported by NSF grant AST-181440. TAT is supported in part by NASA grant 80NSSC20K0531.

Parts of this research were supported by the Australian Research Council Centre of Excellence for All Sky Astrophysics in 3 Dimensions (ASTRO 3D), through project number CE170100013.

B.S.G. was supported by Thomas Jefferson Chair for Discovery and Space Exploration at the Ohio State University.

DEMONEXT was supported by National Science Foundation CAREER Grant AST-1056524 and the Vanderbilt Initiative in Data-intensive Astrophysics (VIDA).

This work is based on observations made by ASAS-SN and ATLAS. The authors wish to recognize and acknowledge the very significant cultural role and reverence that the summit of Haleakal\=a has always had within the indigenous Hawaiian community.  We are most fortunate to have the opportunity to conduct observations from this mountain.

\bibliography{bibliography}
\bibliographystyle{mnras}


\label{lastpage}
\end{document}